\documentclass[%
 aip,
 apl,
 amsmath,amssymb,
 reprint,%
]{revtex4-1}

\usepackage[utf8]{inputenc}
\usepackage[T1]{fontenc}
\usepackage[english]{babel}

\usepackage{lmodern}

\usepackage{mathtools}
\usepackage{amssymb} 

\usepackage{bm} 

\usepackage{pifont}

\usepackage[version=3]{mhchem} 
\usepackage{esint} 
\usepackage{dsfont} 
\usepackage[makeroom]{cancel} 
\usepackage{empheq}
\usepackage{physics}
\usepackage[binary-units=true]{siunitx}

\usepackage{graphicx}

\usepackage{booktabs}
\usepackage{makecell}
\usepackage{threeparttable}
\usepackage{multirow}
 
\usepackage{enumitem}

\usepackage[dvipsnames]{xcolor}

\usepackage{varioref} 
\PassOptionsToPackage{hyphens}{url}
\usepackage[hidelinks,breaklinks=true,linktoc=all]{hyperref} 
\hypersetup{colorlinks=true,
allcolors=black} 
\usepackage[nameinlink,capitalise,noabbrev]{cleveref} 
\crefformat{equation}{(#2#1#3)}
\Crefformat{equation}{Equation (#2#1#3)}
\crefrangeformat{equation}{(#3#1#4) to~(#5#2#6)}
\crefmultiformat{equation}{(#2#1#3)}%
{ and~(#2#1#3)}{, (#2#1#3)}{ and~(#2#1#3)}
\Crefmultiformat{equation}{Equations (#2#1#3)}%
{ and~(#2#1#3)}{, (#2#1#3)}{ and~(#2#1#3)}

\crefname{appendix}{appendix}{appendices}
\Crefname{appendix}{Appendix}{Appendices}

\usepackage{etoolbox}
\usepackage{pgf,tikz,pgfplots}
\pgfplotsset{compat=1.10}
\usetikzlibrary{shapes,arrows,calc,arrows.meta,
patterns,backgrounds,positioning,fit}
\tikzset{every picture/.style=black}
\usepackage{circuitikz}

\tikzset{three sided/.style={
        draw=none,
        append after command={
            [shorten <= -0.5\pgflinewidth]
            ([shift={(-1.5\pgflinewidth,-0.5\pgflinewidth)}]				  \tikzlastnode.north east)
        edge([shift={( 0.5\pgflinewidth,-0.5\pgflinewidth)}]				  \tikzlastnode.north west) 
            ([shift={( 0.5\pgflinewidth,-0.5\pgflinewidth)}]				  \tikzlastnode.north west)
        edge([shift={( 0.5\pgflinewidth,+0.5\pgflinewidth)}]				  \tikzlastnode.south west)            
            ([shift={( 0.5\pgflinewidth,+0.5\pgflinewidth)}]				  \tikzlastnode.south west)
        edge([shift={(-1.0\pgflinewidth,+0.5\pgflinewidth)}]				  \tikzlastnode.south east)
        }
    }
}

\usepackage{psfrag}
\usepackage{pstool} 

\usepackage{lipsum}  

\makeatletter
\newcommand{\getfontsize}{\f@size pt}
\makeatother

\newcommand*\ds{\displaystyle}

\newcommand*{\E}{\mathop{}\!\mathbb{E}}

\newcommand*{\Var}{\mathop{}\!\operatorname{Var}}

\newcommand{\bigO}{\mathcal{O}}

\newcommand*\diff{\mathop{}\!\mathrm{d}}

\newcommand*\kB{k}

\newcommand*\qe{q_{\si{\elementarycharge}}} 

\newcommand*\Vth{V_{\mathrm{th}}}

\newcommand*\VG{V_{\mathrm{G}}}

\newcommand*\VDD{V_{\mathrm{DD}}}

\newcommand*\fmax{f_{\mathrm{max}}}

\newcommand{\dt}{\diff{t}}

\newcommand*\vIN{v_{\mathrm{IN}}}

\renewcommand*\tr{t_{\mathrm{r}}}

\newcommand*\dV{\delta V}

\newcommand*\vC{v_{\mathrm{OUT}}}

\newcommand*\Gav{\overline{G}}
\newcommand*\Dt{\Delta t}
\newcommand*\Dq{\Delta q}
\newcommand*\Dv{\delta v}

\newcommand*\Edissip{\Delta E_{\mathrm{dissip}}}

\newcommand*\Edissipmin{\Delta E_{\mathrm{dissip,min}}}

\newcommand*\dq{\diff{q}}
\newcommand*\dEdissip{\diff{E_{\mathrm{dissip}}}}

\definecolor{k}{rgb}{0 0 0}
\definecolor{r}{rgb}{1 0 0}
\definecolor{g}{rgb}{0 1 0}
\definecolor{b}{rgb}{0 0 1}
\definecolor{orange}{rgb}{1,0.7,0}
\definecolor{c}{rgb}{0 1 1}
\definecolor{cc}{RGB}{64 224 208}
\definecolor{m}{rgb}{1 0 1}
\definecolor{khaki}{RGB}{128 128 0}
\definecolor{deepskyblue}{RGB}{0 191 255}
\definecolor{darkMagenta}{rgb}{0.5 0 0.5}
\definecolor{chocolateBrown}{RGB}{98 52 18}
\definecolor{lightBrown}{RGB}{189 154 122}
\definecolor{mybrown}{RGB}{127 37 0}
\definecolor{bordeaux}{RGB}{131 41 85}
\definecolor{myGreen}{RGB}{134,180,44}
\definecolor{gray_gate}{RGB}{211,208,205}
\definecolor{yellow_oxide}{RGB}{244,231,164}

\definecolor{h}{rgb}{0 0 0}

\definecolor{l}{rgb}{0 0 0}

\definecolor{hh}{rgb}{0 0 0}

\graphicspath{{figures/}}

\makeatletter
\def\@email#1#2{%
 \endgroup
 \patchcmd{\titleblock@produce}
  {\frontmatter@RRAPformat}
  {\frontmatter@RRAPformat{\produce@RRAP{*#1\href{mailto:#2}{#2}}}\frontmatter@RRAPformat}
  {}{}
}%
\makeatother
\begin{document}

\preprint{AIP/123-QED}

\title[Noise-Dissipation Relation for Nonlinear Electronic Circuits]{Noise-Dissipation Relation for Nonlinear Electronic Circuits}

\author{Léopold Van Brandt}
\email{leopold.vanbrandt@uclouvain.be}
\author{Jean-Charles Delvenne}%
\affiliation{ 
        Institute for Information and Communication Technologies,\\ Electronics and Applied Mathematics (ICTEAM)\\
        UCLouvain, Louvain-la-Neuve, Belgium\\
}


\date{\today}

\begin{abstract}
An extension of fluctuation-dissipation theorem is used to derive a `speed limit' theorem for nonlinear electronic devices. This speed limit provides a lower bound on the dissipation that is incurred when transferring a given amount of electric charge in a certain amount of time with a certain noise level (average variance of the current). This bound, which implies a high energy dissipation for fast, low-noise operations (such as switching a bit in a digital memory), brings together recent results of stochastic thermodynamics into a form that is usable for practical nonlinear electronic circuits, 
as we illustrate on a switching circuit made of an nMOS pass gate in a state-of-the-art industrial technology. 
\end{abstract}

\maketitle

Electronic circuits operate at an energetic cost ---under the form of dissipation in resistive parts of the circuit--- that brings technological or economical costs (need for cooling, large batteries, etc.), as well as ecological issues \cite{Bol2015}. \textcolor{hh}{The search for more energy-efficient devices} sometimes runs into the problem of reliability, as the intrinsic noise level may be become non-negligible at low voltages \cite{EUROSOI2019,Freitas2021stochastic,Freitas2022reliability,EUROSOI2023}.
In parallel, there is a quest for theoretical lower bounds on dissipated energy and noise valid for a range of physical 
systems (electronic or else) fulfilling certain tasks, notably related to computation. One early such bound is Landauer's bound \cite{Landauer1961}, that states that erasing a bit necessarily dissipates an energy $\kB T \ln 2$, in whatever technology.

Stochastic thermodynamics \cite{Peliti2021} offers a recent framework where bounds beyond Landauer's can be formulated and proved rigorously for broad classes of physical systems. Such bounds include the Thermodynamic Uncertainty Relations \cite{Horowitz2020,Falasco2020}, which state that stationary systems exhibiting a low level of noise must necessarily dissipate a lot --- in other words, suppressing noise is costly. Another family of bounds, the (Classical) Speed Limits \cite{Shiraishi2018}, provide specific trade-offs between the time of an operation (e.g.  writing a bit into a memory) and its dissipation, quantifying precisely the well-known observation that fast operations cost more.

The existing Speed Limits are not always straightforward to use for electronic circuits, as they rely on assumptions on the form of noise (e.g. discrete jumps) not always satisfied in practice. A recent Speed Limit \cite{IEEENANO2023} is naturally applicable to electronic circuits, but does not take noise into account, missing an essential component of the dissipation-time-noise trade-off. 

We propose such a trade-off applicable to any linear or nonlinear resistive device, regardless of the noise model. 
It is a relationship between the total average charge passing through a nonlinear resistive device over a time interval, the energy dissipated in the device, and the total noise (variance) over that interval.

\newcommand\myfontsize{\normalsize}
\newcommand\dx{0.75}
\newcommand\xspacing{1.5}
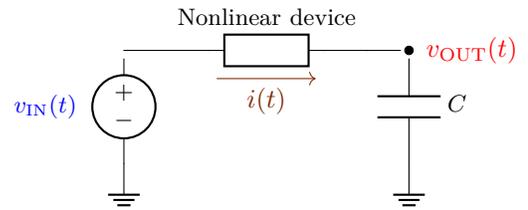
\begin{figure}
\centering
\begin{circuitikz}[european, american voltages, transform shape, line cap=rect, nodes={line cap=butt},scale=1]
\draw[draw=none,opacity=0]
(0,0) node[nmos,rotate = -90,yscale=-1] (nMOS) {};
\draw
(nMOS.D) to[R=Nonlinear device,-] (nMOS.S);
\draw 
(nMOS.D) -- ++(-\xspacing*\dx,0) node[] (IN) {};
\draw 
(IN) ++ (0,-2*\dx) node[ground,scale=1,color=black] (GNDIN) {}
to[V=\textcolor{b}{$\vIN(t)$},invert] (IN);
\draw 
(nMOS.S) ++(\xspacing*\dx,0) node[label={[font=\myfontsize]right:$\textcolor{r}{\vC(t)}$}] (OUT) {}
(nMOS.S) to[short,-*] (OUT);
\draw
(OUT)++(0,-2*\dx) node[ground,scale=1,color=black] {} to[C,color=black,l_={$C$}] (OUT);
{
\color{mybrown}
\ctikzset{current/distance = 0.5}
\draw 
(nMOS.D)++(0,-0.5*\dx) node[] (nDi) {}
(nMOS.S)++(0,-0.5*\dx) node[] (nSi) {};
\draw[->] 
(nDi) -- node[below, color=mybrown, font=\myfontsize] {$i(t)$} (nSi);
}
\end{circuitikz}
\caption{Charging a capacitance through some (possibly nonlinear) dissipative noisy device.}
\label{fig_statement}
\end{figure}

 We suppose that the nonlinear device is purely resistive with negligible internal dynamics, but is otherwise  arbitrary. The device is embedded into an arbitrary circuit. 
For illustrative purposes and without loss of generality, we assume the simple circuit represented in \cref{fig_statement}. This circuit may, for instance, be interpreted as writing a bit from logical 0 to logical 1 by transferring a certain amount of charge into the capacitance.

A (nonlinear) resistive device at temperature $T$ is intrinsically noisy \textcolor{hh}{due to the random agitation of the charge carriers (electrons)}. Indeed, when 
subjected
to a constant voltage difference $\Dv$ \textcolor{h}{applied externally} 
during a
time $\dt$, it is traversed by a \emph{random} charge $\Dq$, dissipating on the way an energy $\Edissip=\Dq \, \Dv$. \textcolor{hh}{Charges during distinct time intervals are statistically independent, i.e. the noise on current is white (possibly Gaussian, or Poisson, or else).}

For a linear  device, mean and variance of $\Dq$ are 
related
by the celebrated fluctuation-dissipation 
theorem \cite{Kubo1966},
which translates into the also famous Johnson-Nyquist formula \cite{Johnson1928,Nyquist1928} for a linear resistor\textcolor{h}{, holding for all equilibrium ($\Dv = 0$) and non-equilibrium conditions \cite{Wyatt1999}}.
Arbitrary nonlinear devices follow a more general relationship linking all moments, the fluctuation relation \cite{Jarzynski1997}, a consequence of which is the Thermodynamic Uncertainty Relation~\cite{Horowitz2020,IFAC2023}:
\begin{equation}
\label{eq:TUR}
\frac{\E\{\Edissip\}}{kT} \geq 2 \, \frac{\E\{\textcolor{h}{\Dq}\}^2}{\Var \{\Dq\}}
\text{.}
\end{equation}
In \eqref{eq:TUR}, both the mean $\E\{\Dq\}$ and the variance $\Var \{\Dq\}$ are proportional 
\textcolor{hh}{to $\Dt$ (due to whiteness of the noise),
hence both sides of \eqref{eq:TUR} scale} as $\propto \Dt$.
It was shown that the relation is tight (i.e. is an equality) if the random fluctuations are symmetric around the mean, e.g.  for Gaussian fluctuations \cite{IFAC2023}.  

When the voltage across the device is time-varying (non-stationary conditions, for instance in a switching digital circuit) and random, we may still apply \eqref{eq:TUR} to a small or infinitesimal interval $\dt$ within the whole time interval $[t_0,t_0+\Dt]$, and conditionally to a given value $\Dv(t)$. This leads naturally to, for any voltage difference $\Dv(t)=\Dv$ over any infinitesimal time interval $[t,t+\dt]$:
\begin{equation}
\label{eq:TURcond}
\frac{\E \{\dEdissip|\Dv(t)=\Dv\}}{kT} \geq 2 \, \frac{\E\{\dq|\Dv(t)=\Dv\}^2}{\Var \{\dq|\Dv(t)=\Dv\}}
\text{.}
\end{equation}
Because 
we are interested in the unconditional mean dissipation 
$\E \{\dEdissip \}=	\E_{\Dv} \E\{\dEdissip|\Dv(t)=\Dv\}$ 
(where $\E_{\Dv}$ refers to averaging over all values of $\Dv$), we may write
\begin{equation}
\label{eq:TURcond 2}
\frac{\E\{\dEdissip\}}{kT} \geq 2 \, \E_{\Dv} \Big\{ \frac{\E\{\dq|\Dv(t)=\Dv\}^2}{\Var \{\dq|\Dv(t)=\Dv\}} \Big\}
\text{.}
\end{equation}
\textcolor{h}{However}, we would like to obtain an expression involving unconditional means and variances, more accessible and interpretable than their conditional variants.

To that purpose, we can define a scalar product between arbitrary (square integrable) real-valued functions 
$f(\Dv)$ and $g(\Dv)$
\begin{equation}
\langle f|g \rangle=\E_{\Dv} \{ f(\Dv)g(\Dv) \}
\end{equation}

As all scalar products, it satisfies the well known Cauchy-Schwarz inequality $\langle f|f \rangle \geq \langle f|g \rangle^2 / \langle g|g \rangle$. 
Let us define $f\equiv\E\{\dq|\Dv(t)=\Dv\}/\sqrt{\Var\{\dq|\Dv(t)=\Dv\}}$ and $g \equiv \sqrt{\Var\{\dq|\Dv(t)=\Dv\}}$, and apply Cauchy-Schwarz inequality.
We recognize $\langle f|f \rangle$ as the r.h.s. of \eqref{eq:TURcond 2}, up to the factor $2$.
Furthermore, we observe from the law of total variance that
\[
\begin{aligned}
\Var\{\dq\} 
&= \E_{\Dv} \Var\{\dq|\Dv(t)=\Dv\} + \Var_{\Dv} \E\{\dq|\Dv(t)=\Dv\} \\
& =\E_{\Dv} \Var\{\dq|\Dv(t)=\Dv\}  + \bigO\big(\dt^2\big)
\text{,}
\end{aligned}
\]
the latter term being thus negligible.
We finally get:
\begin{equation}
\label{eq:TUR small}
\frac{\E\{\dEdissip\}}{kT} \geq 2 \, \frac{\E\{\dq\}^2}{\Var \{\dq\}}.
\end{equation}
Integrating \eqref{eq:TUR small} over the whole time interval $[t_0,t_0+\Dt]$, we obtain a lower bound on dissipation, which is our first main result:
\begin{equation}
\label{eq:TUR int}
\frac{\E\{\Edissip\}}{kT}
\geq
 2\, \int_{t_0}^{t_0 + \Dt} \! \frac{\E\{\dq\}^2}{\Var \{\dq\}}
\text{.}
\end{equation}
In \eqref{eq:TUR int}, $\E\{\Edissip\}$ is the average \emph{energy dissipation} over the whole time interval $[t_0,t_0+\Dt]$, i.e.
\begin{equation}
\label{eq:Edissip}
\E\{\Edissip\} 
= \int_{t_0}^{t_0 \mathrlap{+ \Dt}} \E\{\dEdissip\}
= \int_{t_0}^{t_0 \mathrlap{+ \Dt}} \E\{\dq(t)\,  \Dv(t)\} \, \dt
\text{.}
\end{equation} 

We can relax further this inequality by applying again the Cauchy-Schwarz inequality. 
We now use the scalar product $\langle f|g \rangle'=\int_{t_0}^{t_0+\Dt} f' g' \dt$ on real-valued functions 
$f'(t)$ and $g'(t)$
of time over the interval $[t_0,t_0+\Delta t]$. We apply it to $f'(t)\equiv\E\{\dq(t)\}  / \sqrt{\Var\{\dq(t)\}}$ and $g'(t) \equiv \sqrt{\Var\{\dq(t)\}}$ to obtain
\begin{equation}
\label{eq:TUR2}
\frac{\E\{\Edissip\}}{kT} 
\geq 2 \, 
\frac{\ds \E\{\Dq\}^2}{\ds \int_{t_0}^{t_0+\Dt} \! \Var \{\dq\}}=2 \, 
\frac{\ds \E\{\Dq\}^2}{\ds \overline{\Var} \{\dq\} \,  \Dt}
\text{,}
\end{equation}
which is our second main result. The time-averaged variance  $\overline{\Var}\{\dq\}$ is $\int_{t_0}^{t_0+\Dt} \Var \{\dq\}/\Dt$.
Although
this bound is not as tight as \eqref{eq:TUR int}, it may prove easier to evaluate and interpret in many cases. In particular, $\Dq=\int_{t_0}^{t_0+\Dt} \dq$ in the numerator is the total charge passed through the device over the interval, which is usually the quantity of interest. 

{\color{hh}
In summary, for a given $\Dq$, fast charge transfer (low $\Dt$) and/or low-noise process (small $\overline{\Var} \{\dq\}$) implies large dissipation.
}
The relationship \eqref{eq:TUR2} can indeed be seen as a novel \emph{speed limit} relation: passing a certain charge $\Dq$ over a device with a typical level of noise $\overline{\Var}\{\dq\}$ within a duration $\Dt$ necessarily dissipates an energy that is inversely proportional to $\Dt$.  This speed limit differs from other recent speed limits recently obtained that also obtain a $1/\Dt$ behaviour. For instance, most speed limits \cite{Falasco2022} only apply to discrete jumps of charges (e.g. shot noise), while we cover all noise sources, discrete or continuous with a single formula. 
Let us also mention the recent deterministic speed limit~\cite{IEEENANO2023},
\begin{equation}
\label{eq:Edissipmin}
\Edissip \geq \Edissipmin 
\equiv \frac{\Dq^2}{\Gav \, \Dt}
\text{,}
\end{equation}
%
%
 which does not directly take noise into account. In \eqref{eq:Edissipmin}, $\Gav$ denotes the time-averaged conductance 
of the driving device.
  
Finally, we also see that for a given $\Dt$, a low noise level (low variance) can only be obtained at the cost of high dissipation. 
Our bounds \cref{eq:TUR2,eq:TUR int} thus express a trade-off between speed, dissipation and noise 
for an arbitrary nonlinear resistive device.

Our preliminary application is the case where the driving device in \cref{fig_statement} is a linear resistor of conductance $G$, that can be covered in detail analytically.
From Ohm's law,  $\E\{dq(t)|\dV(t)=\dV\} = G \dV \dt$. Johnson-Nyquist's formula~\cite{Johnson1928,Nyquist1928} is an electrical equivalent of Einstein's diffusion law \cite{Einstein1905}, and an avatar of the general fluctuation-dissipation theorem~\cite{Kubo1966}, stating here that $\Var \dq(t)=2kT G dt$.

Note in passing that Johnson-Nyquist's formula is often expressed in the electronic literature for the stationary random current $i(t)=dq/dt$ for a constant $\dV$. This current is  defined (as a random function of time) only in a weak sense, i.e. if we limit it to a finite frequency bandwidth $\Delta f$. The variance is then $\Var \{i(t)\}= 4kTG \Delta f$. In this case we can interpret the current defined by $i(t)=dq(t)/dt$ (for a small, non-infinitesimal $\dt$) as being limited to the frequency band $[0,1/2\dt]$.  

The system is thus a linear system with an external driving $v_{IN}(t)$ and an internal noise, described by a linear stochastic differential equation (a Langevin equation more precisely), which can be solved explicitly. 
In this context, \eqref{eq:TURcond} is satisfied with equality, and our main result \eqref{eq:TUR2} becomes:
\begin{equation}
	\label{eq:Edissipmin RC}
	\Edissip \geq \Edissipmin =  \frac{\E\{\Dq\}^2}{G\Dt}
	\text{,}
\end{equation}
which is precisely \eqref{eq:Edissipmin}, the minimum dissipation from the deterministic speed limit\cite{IEEENANO2023}, for a constant capacitance load and by identifying $\Gav = G$. In the linear case, thanks to Johnson-Nyquist's noise model (a specific case of the fluctuation-dissipation theorem) turning \cref{eq:TUR,eq:TURcond} into equalities, our bound coincides with a deterministic treatment\cite{IEEENANO2023}. We shall now see it is not the case in a nonlinear case.



\begin{figure}
\centering
\begin{circuitikz}[american voltages, transform shape, line cap=rect, nodes={line cap=butt},scale=0.9]
\draw[black]
(0,0) node[nmos,rotate = -90,yscale=-1] (nMOS) {}
(nMOS.D) to[short,-o] ++(0,0) node[label={[font=\myfontsize,color=black]above:D}] (nD) {}
(nMOS.G) to[short,-o] ++(0,0) node[label={[font=\myfontsize,color=black,yshift=-0.1cm]above:$\VG = \VDD$}] (nG) {}
(nMOS.S) to[short,-o] ++(0,0) node[label={[font=\myfontsize,color=black]above:S}] (nS) {}
;
\draw 
(nMOS.D) -- ++(-\xspacing*\dx,0) node[] (IN) {};
\draw 
(IN) ++ (0,-2*\dx) node[ground,scale=1,color=black] (GNDIN) {}
to[V=\textcolor{b}{$\vIN(t)$},invert] (IN);
\draw 
(nMOS.S) ++(\xspacing*\dx,0) node[label={[font=\myfontsize]above:\textcolor{r}{$\vC(t)$}}] (OUT) {}
(nMOS.S) to[short,-*] (OUT);
\draw
(OUT)++(0,-2*\dx) node[ground,scale=1,color=black] {} to[C,color=black,l_={$C$}] (OUT);
{
\color{mybrown}
\ctikzset{current/distance = 0.5}
\draw 
(nMOS.D)++(0,-0.3*\dx) node[] (nDi) {}
(nMOS.S)++(0,-0.3*\dx) node[] (nSi) {};
\draw[->] 
(nDi) -- node[below, color=mybrown, font=\myfontsize] {$i(t)$} (nSi);
}
\end{circuitikz}
\caption{nMOS pass gate driving a capacitor.}
\label{fig_pass_gate}
\vspace{2mm}
\centering
(a)
\vspace{-1mm}
\psfragscanon
\footnotesize
\psfrag{v(t) [V]}[cc][cc]{\normalsize  $v(t) \, [\si{\volt}]$}
\psfrag{0}[cc][cc]{$0$}
\psfrag{0.5}[cc][cc]{$0.5$}
\psfrag{0.8}[cc][cc]{$0.8$}
\psfrag{1}[cc][cc]{$1$}
\psfrag{vIN(t)}[tl][tl]{\small $\textcolor{b}{\vIN(t)}$}
\psfrag{vOUT(t)}[tl][tl]{\small A sample path $\textcolor{r}{\vC(t)}$}
\psfrag{a.VDD}[bl][bl]{\small $V_1 = \alpha \, \VDD$}
\psfrag{tr}[tc][tc]{\small  $\textcolor{b}{\tr}$}
\psfrag{I}[bc][bc]{I}
\psfrag{II}[bc][bc]{II}
\psfrag{Dt}[cl][cl]{$\Dt$}
\includegraphics[scale=1]{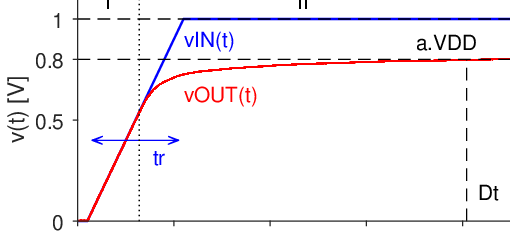}
(b)
\vspace{-1mm}
\psfragscanon
\footnotesize
\psfrag{i(t) [uA]}[cc][cc]{\normalsize $i(t) \, [\si{\micro\ampere}]$}
\psfrag{-1.5}[cc][cc]{$-1.5$}
\psfrag{-1}[cc][cc]{$-1$}
\psfrag{-0.5}[cc][cc]{$-0.5$}
\psfrag{0}[cc][cc]{$0$}
\psfrag{0.5}[cc][cc]{$0.5$}
\psfrag{1}[cc][cc]{$1$}
\psfrag{1.5}[cc][cc]{$1.5$}
\psfrag{i(t)}[cl][cl]{\small A sample path $\textcolor{mybrown}{i(t)}$}
\psfrag{E(i(t))}[cl][cl]{\small Empirical $\textcolor{orange}{\E\big\{i(t)\big\}}$ (from 1000 
paths)}
\includegraphics[scale=1]{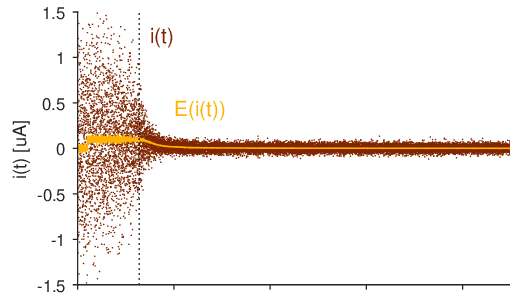}
\vspace{2mm}
(c)
\psfragscanon
\footnotesize
\psfrag{t [ns]}[cc][cc]{\raisebox{+2mm}{\normalsize $t \, [\si{\nano\second}]$}}
\psfrag{(E(dq))^2, Var(dq) [qe^2]}[cc][cc]{\small \textcolor{deepskyblue}{$\big(\E\{\dq(t)\}\big)^2$}, \textcolor{myGreen}{$\Var\{\dq(t)\}$} $[\qe^2]$}
\psfrag{0}[cc][cc]{$0$}
\psfrag{10}[cc][cc]{$10$}
\psfrag{20}[cc][cc]{$20$}
\psfrag{30}[cc][cc]{$30$}
\psfrag{40}[cc][cc]{$40$}
\psfrag{50}[cc][cc]{$50$}
\psfrag{60}[cc][cc]{$60$}
\psfrag{80}[cc][cc]{$80$}
\psfrag{100}[cc][cc]{$100$}
\psfrag{0.5}[cc][cc]{\color{darkMagenta}$0.5$}
\psfrag{1}[cc][cc]{\color{darkMagenta}$1$}
\psfrag{E(dv(t)i(t))dt}[cl][cl]{\small\color{gray} \hspace{-2mm}$\ds\E\big\{(\vIN(t)-\vC(t))\,  i(t) \big\} \frac{\dt}{2 \, \kB T}  \ \rightarrow$}
\psfrag{ratio}[cl][cl]{\small\color{darkMagenta} \hspace{-2mm}$\ds\frac{\E\{\dq\}^2}{\Var \{\dq\}} \ \rightarrow$}
\includegraphics[scale=1]{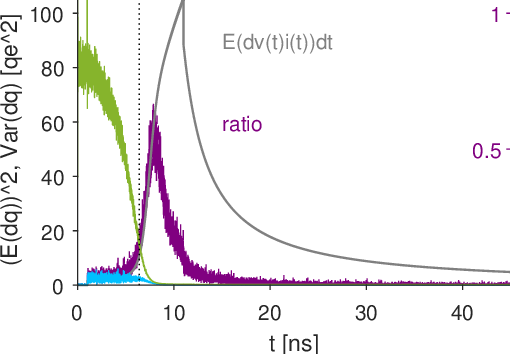}
\vspace{-6mm}
\caption{
Transient noise simulation of the bit writing. 
\newline
Circuit parameters: $\SI{28}{\nano\meter}$ FD-SOI regular-$\Vth$ nMOS of minimal dimensions $L = \SI{30}{\nano\meter}$ and $W = \SI{80}{\nano\meter}$;
$T = \SI{25}{\celsius}$;
$\VDD = \SI{1}{\volt}$;
$C = \SI{1}{\femto\farad}$;
input rise time $\tr = \SI{10}{\nano\second}$.
\newline
Bandwidth of 
generated noise: $\fmax = \SI{200}{\giga\hertz}$; $\dt = \SI{2.5}{\pico\second}$.
\newline
(a) voltage signals; (b) current trace; (c) statistics of $\dq(t)$ empirically \emph{estimated} from 1000 sample paths (left); normalized dissipation versus ratio 
$\E\{\dq\}^2/\Var \{\dq\}$
of \eqref{eq:TUR int} (right).}
\label{fig_pass_gate_tr_10ns_NOISETRAN}
\end{figure}

Because the scope and the significance of our main theoretical results \cref{eq:TUR2,eq:TUR int} are only fully appreciated in 
\textcolor{hh}{circuits} involving a nonlinear device, we illustrate their application to the circuit of \cref{fig_pass_gate}: 
a constant capacitor is charged through a pass gate, implemented with a MOS (Metal-Oxide Semiconductor) transistor.
The simulated signals and extracted quantities are depicted in \cref{fig_pass_gate_tr_10ns_NOISETRAN}.
The charging process is, without loss of generality, here ensured by the linear input voltage ramp of amplitude $\VDD = \SI{1}{\volt}$ sketched in \cref{fig_pass_gate_tr_10ns_NOISETRAN}(a), corresponding to the supply voltage of the used CMOS technology. 
The bit writing operation is assumed finished when $\vC(t)$ reaches $V_1 = \SI{80}{\percent} \, \VDD$ (shown in (\cref{fig_pass_gate_tr_10ns_NOISETRAN}(a)).

In the most advanced technologies favoured for digital circuit design \cite{Baker}, the noise model of the transistor can be very complex (since, in the charge-based modelling approach, it depends on the small-dimension effects part of the deterministic static model \cite{Tsividis2011}) and not known in an insightful closed form.
We therefore avoided to resort to a simplified and inaccurate analytical model of the transistor and we have performed SPICE noise simulation in the time domain (\cref{fig_pass_gate_tr_10ns_NOISETRAN}), compatible with the process design kit provided by the semiconductor foundry \cite{Poiroux2013}. 

Industrial simulators, such as Eldo\textsuperscript{\textregistered}, 
provide a transient noise analysis tool \cite{EUROSOI2023}.
All noise sources are modelled as Gaussian, irrespective of existing discussions of the physics origin of the thermal noise and the thermodynamic inconsistency of the pure Gaussian model \cite{Wyatt1999,IFAC2023}.
Individual noise source are dynamically generated in the time domain according the compact model of the device, and the noise is then handled as any other electrical signal during a transient simulation.

The intrinsic noise is the noise generated by the MOS transistor of \cref{fig_pass_gate}.
Within SPICE formalism, the intrinsic noise of the MOS transistor, that is the noise produced by the device itself, is modelled as a current \textcolor{h}{Gaussian} fluctuation.
One sample path of $i(t)$ (among $1000$ generated and processed) is represented in \cref{fig_pass_gate_tr_10ns_NOISETRAN}(b), as well as the empirical mean.
Whereas the physical white noise is, 
neglecting quantum effects, 
of infinite bandwidth, 
the noise is generated within a specified finite bandwidth, denoted $\fmax$, for the transient simulation.
The $\fmax$ must be selected in order to capture the dominant effect of the thermal noise on the charge transfer,
extracted as about $\SI{200}{\giga\hertz}$, beyond the bandwidth of the circuit of \cref{fig_pass_gate} for different $\vIN$ from $0$ to $\SI{1}{\volt}$.

We have recorded 1000 sample paths (or \emph{trajectories)} or \emph{traces}) of the voltage and current signals simulated in the time domain.
A specified $\fmax$ imposes a constraint on the (maximum) simulation time step ($\dt$) to be used~\cite{EUROSOI2023}: $\dt = 1/(2\,\fmax)$, according to Shannon-Nyquist sampling theorem.
The 
time step $\dt = \SI{2.5}{\pico\second}$ is small enough 
to be numerically assimilated to an infinitesimal time interval, both to compute the charge increment
\begin{equation*}
\dq(t) = i(t)\,\dt = C\,(\vC(t+\dt) - \vC(t))
\end{equation*}
and the different integral quantities.
The sample mean and variance (of $i(t)$ in \cref{fig_pass_gate_tr_10ns_NOISETRAN}(b) and of $\dq(t)$ in \cref{fig_pass_gate_tr_10ns_NOISETRAN}(c)) are unbiased empirical estimators \cite{Papoulis1991}, and the variability observed for these statistics in \cref{fig_pass_gate_tr_10ns_NOISETRAN}(b,c) could of course be reduced at the expense of an increased number of simulated paths.

In \cref{fig_pass_gate_tr_10ns_NOISETRAN}(c), we first depict the empirical estimators of the two important quantities of  \eqref{eq:TUR int}, $\E\{\dq(t)\}^2$ and $\Var \{\dq(t)\}$ (left $y$ axis).
Generally, we expect from the relations \cref{eq:TUR2,eq:TUR int} that the ratio $\E\{\dq\}^2/\Var \{\dq\}$ 
predicts the trend of the evolution over time of the dissipation, 
i.e. that the dissipation is locally large when the intrinsic noise fluctuations (relative to the average instantaneous charge transfer) is low (and conversely).
The discrepancy over time between the actual energy dissipation and the ratio $\E\{\dq\}^2/\Var \{\dq\}$ (right $y$ axis) tells us how tight we can expect the lower bounds \cref{eq:TUR2,eq:TUR int} to be (see coming results), and thereby assesses the energy efficiency of the charge transfer through the nonlinear dissipative device (of some intrinsic noise level) for the considered input signal and speed. 

By inspecting first \cref{fig_pass_gate_tr_10ns_NOISETRAN}(a), {\color{h}
we can basically distinguish two regimes of the different energy efficiency for the charge transfer, 
characterized by a larger or lower conductance of the nonlinear driving device. For the (important) special case of the MOS transistor (\cref{fig_pass_gate}), the regime may be referred to as \emph{inversion level} (or \emph{region}) \cite{Tsividis2011}.
}
In the first (I) regime, the charging process is efficient in the sense that $\vC(t)$ closely follows  $\vIN(t)$ (\textcolor{hh}{the so called quasi-adiabatic conditions \cite{Snider2012,Orlov2012}})
and the dissipation is low.
{\color{hh}
For the case illustrated in \cref{fig_pass_gate_tr_10ns_NOISETRAN}, the speed of the charging process, $\dq/\dt$, is low and constant during regime I, which simplifies the interpretation of the general results \cref{eq:TUR2,eq:TUR int}.
}
The transistor is in strong inversion, meaning that it is highly conductive and also that the intrinsic current noise is large \cite{Tsividis2011}.
This is indeed what we observe in \cref{fig_pass_gate_tr_10ns_NOISETRAN}(b,c): $\Var \{i(t)\}$ (or $\Var \{\dq(t)\}$) is large and drops gradually (while $\E\{\dq(t)\}^2$ does not vary much).
In the second (II) regime, the transistor falls in weaker inversion (low conductance) and $\vC(t)$ painfully follows $\vIN(t)$ \cite{IEEENANO2023} (“the nMOS is not good at passing a 1” \cite{Baker}).
The dissipation is large (see the gray shark-fin-shaped peak
in \cref{fig_pass_gate_tr_10ns_NOISETRAN}(c)) \textcolor{hh}{as evidenced by the high-amplitude intrinsic noise fluctuations in this regime}.
We believe that this circuit application highlights the existing correlation between fluctuation and dissipation trend, yet in dynamic and nonlinear conditions.
Finally, let us emphasize that, in \cref{fig_pass_gate_tr_10ns_NOISETRAN}(c), the discrepancy between actual dissipation and $\E\{\dq\}^2/\Var \{\dq\}$ becomes significant after $\SI{10}{\nano\second}$, precisely when the charge transfer gets even more inefficient.

\begin{table}[]
\small
\newcommand*{\tabindent}{\hspace{1cm}}
\newcommand*{\tabjump}{\vspace{0.25cm}}
\centering
\begin{tabular}{lc}
\toprule
Energy dissipation & Value $[\kB T]$
\\
\midrule
Actual $\E\{\Edissip\}$ & 
\\
\tabindent 
	Eq.~\eqref{eq:Edissip} & \num{6919} 
\\
\midrule
\tabjump 
Speed limit bounds of this article &
\\
\tabindent 
	Integral Eq.~\eqref{eq:TUR int} & \num{1568} 
\tabjump \\
\tabindent Eq.~\eqref{eq:TUR2} over $[t_0, t_0 + \Dt]$ & \num{375}
	\\
\tabjump \\
Deterministic speed limit \cite{IEEENANO2023} &
\\
\tabindent Eq.~\eqref{eq:Edissipmin} & \num{213}
\\
\bottomrule
\end{tabular}
\caption{
Actual energy dissipation compared to the different lower bounds derived or referred in this paper. All the quantities were extracted from the simulation of \cref{fig_pass_gate_tr_10ns_NOISETRAN}\textcolor{h}{, notably $\Dt \approx \SI{40}{\nano\second}$ defined at $V_1 = \SI{80}{\percent} \, \VDD$}.
}
\label{tab:Edissip}
\end{table}

In \cref{tab:Edissip}, we summarize the dissipation-related quantities extracted from the transient simulation presented in \cref{fig_pass_gate_tr_10ns_NOISETRAN}.
The expected or average dissipation, computed according to the definition \eqref{eq:Edissip} is the reference value to which we compare our different lower bound.
These are listed in ascending order in \cref{tab:Edissip}, which is consistent with their order of appearance in the text.

The integral bound \eqref{eq:TUR int} is lower than the actual dissipation only by a factor \num{4.4}.
At the light of \cref{fig_pass_gate_tr_10ns_NOISETRAN}(c), we have attributed such discrepancy to the second regime of the charging process, where the nMOS transistor becomes highly inefficient in fully passing a logical 1. The worsened conductivity (larger dissipation) is correlated with a lower noise level, reality reflected in the relationship \eqref{eq:TUR int}.

Importantly, the reported discrepancy reveals the non-Johnson-Nyquist nature of the noise fluctuation-dissipation process of the nonlinear device \cite{IFAC2023} (as opposed to the linear resistor).
We know that the shot noise model is consistent with a MOS transistor operation in weak inversion \cite{Tsividis2011}. 
In case of shot noise\cite{IFAC2023}, the relation \eqref{eq:TUR} is loose by a factor of \num{4} for a voltage difference $\dV \approx 200 mV$ (i.e., about $\num{8}$ times the \textcolor{h}{thermal voltage} $\kB T/\qe$, see Figure 1 in the reference\cite{IFAC2023}), which is broadly consistent with our numerical observations.  This shows that the fact that even though our bound holds for any model of noise, different models will make the bound more or less tight.

The lower bound \eqref{eq:TUR2} is, as announced earlier and here verified experimentally, less tight than \eqref{eq:TUR int} as lying more than one order of magnitude below the dissipation.



To conclude, we have exploited the theoretical framework of the stochastic thermodynamics to propose noise-dissipation relations valid for in non-equilibrium and non-stationary conditions, relevant for switching digital circuits that are strongly nonlinear.
Two different lower bounds were provided for the energy dissipation. The relations involves the time-varying statistics of the noise over the charge transfer process.
We have applied and discuss them over linear and nonlinear dipoles.
The quantitative analyses deduced from the simulations are insightful about the physical origin of the noise, that cannot merely be pure Johnson-Nyquist noise for nonlinear dissipative devices \cite{IFAC2023}.
Further work would deepen this aspect, in link with recent noise measurement and modelling work \cite{EUROSOI2019,EUROSOI2023}, and would extend the mathematical formalism to more complex circuits with multiple dissipative and dynamic components (e.g. CMOS logic gates).

\begin{acknowledgments}
The work has been supported by the Research Project "Thermodynamics of Circuits for Computation" of the National Fund for Scientific Research (F.R.S.-FNRS) of Belgium.

The authors would like to thank Prof. Denis Flandre, Prof. David Bol, Mr. Martin Lefebvre and Mr. Adrian Kneip for the valuable discussions that contributed to this work.
\end{acknowledgments}

\bibliography{bib}

\begin{thebibliography}{25}%
\makeatletter
\providecommand \@ifxundefined [1]{%
 \@ifx{#1\undefined}
}%
\providecommand \@ifnum [1]{%
 \ifnum #1\expandafter \@firstoftwo
 \else \expandafter \@secondoftwo
 \fi
}%
\providecommand \@ifx [1]{%
 \ifx #1\expandafter \@firstoftwo
 \else \expandafter \@secondoftwo
 \fi
}%
\providecommand \natexlab [1]{#1}%
\providecommand \enquote  [1]{``#1''}%
\providecommand \bibnamefont  [1]{#1}%
\providecommand \bibfnamefont [1]{#1}%
\providecommand \citenamefont [1]{#1}%
\providecommand \href@noop [0]{\@secondoftwo}%
\providecommand \href [0]{\begingroup \@sanitize@url \@href}%
\providecommand \@href[1]{\@@startlink{#1}\@@href}%
\providecommand \@@href[1]{\endgroup#1\@@endlink}%
\providecommand \@sanitize@url [0]{\catcode `\\12\catcode `\$12\catcode
  `\&12\catcode `\#12\catcode `\^12\catcode `\_12\catcode `\%12\relax}%
\providecommand \@@startlink[1]{}%
\providecommand \@@endlink[0]{}%
\providecommand \url  [0]{\begingroup\@sanitize@url \@url }%
\providecommand \@url [1]{\endgroup\@href {#1}{\urlprefix }}%
\providecommand \urlprefix  [0]{URL }%
\providecommand \Eprint [0]{\href }%
\providecommand \doibase [0]{http://dx.doi.org/}%
\providecommand \selectlanguage [0]{\@gobble}%
\providecommand \bibinfo  [0]{\@secondoftwo}%
\providecommand \bibfield  [0]{\@secondoftwo}%
\providecommand \translation [1]{[#1]}%
\providecommand \BibitemOpen [0]{}%
\providecommand \bibitemStop [0]{}%
\providecommand \bibitemNoStop [0]{.\EOS\space}%
\providecommand \EOS [0]{\spacefactor3000\relax}%
\providecommand \BibitemShut  [1]{\csname bibitem#1\endcsname}%
\let\auto@bib@innerbib\@empty
\bibitem [{\citenamefont {Bol}, \citenamefont {de~Streel},\ and\ \citenamefont
  {Flandre}(2015)}]{Bol2015}%
  \BibitemOpen
  \bibfield  {author} {\bibinfo {author} {\bibfnamefont {D.}~\bibnamefont
  {Bol}}, \bibinfo {author} {\bibfnamefont {G.}~\bibnamefont {de~Streel}}, \
  and\ \bibinfo {author} {\bibfnamefont {D.}~\bibnamefont {Flandre}},\ }in\
  \href@noop {} {\emph {\bibinfo {booktitle} {2015 IEEE SOI-3D-Subthreshold
  Microelectronics Technology Unified Conference (S3S)}}}\ (\bibinfo
  {organization} {IEEE},\ \bibinfo {year} {2015})\ pp.\ \bibinfo {pages}
  {1--3}\BibitemShut {NoStop}%
\bibitem [{\citenamefont {Van~Brandt}\ \emph {et~al.}(2019)\citenamefont
  {Van~Brandt}, \citenamefont {Esfeh}, \citenamefont {Kilchytska},\ and\
  \citenamefont {Flandre}}]{EUROSOI2019}%
  \BibitemOpen
  \bibfield  {author} {\bibinfo {author} {\bibfnamefont {L.}~\bibnamefont
  {Van~Brandt}}, \bibinfo {author} {\bibfnamefont {B.~K.}\ \bibnamefont
  {Esfeh}}, \bibinfo {author} {\bibfnamefont {V.}~\bibnamefont {Kilchytska}}, \
  and\ \bibinfo {author} {\bibfnamefont {D.}~\bibnamefont {Flandre}},\ }in\
  \href {\doibase 10.1109/EUROSOI-ULIS45800.2019.9041859} {\emph {\bibinfo
  {booktitle} {2019 Joint International EUROSOI Workshop and International
  Conference on Ultimate Integration on Silicon (EUROSOI-ULIS), Grenoble,
  France}}}\ (\bibinfo {year} {2019})\ pp.\ \bibinfo {pages} {1--4}\BibitemShut
  {NoStop}%
\bibitem [{\citenamefont {Freitas}, \citenamefont {Delvenne},\ and\
  \citenamefont {Esposito}(2021)}]{Freitas2021stochastic}%
  \BibitemOpen
  \bibfield  {author} {\bibinfo {author} {\bibfnamefont {N.}~\bibnamefont
  {Freitas}}, \bibinfo {author} {\bibfnamefont {J.-C.}\ \bibnamefont
  {Delvenne}}, \ and\ \bibinfo {author} {\bibfnamefont {M.}~\bibnamefont
  {Esposito}},\ }\href@noop {} {\bibfield  {journal} {\bibinfo  {journal}
  {Physical Review X}\ }\textbf {\bibinfo {volume} {11}},\ \bibinfo {pages}
  {031064} (\bibinfo {year} {2021})}\BibitemShut {NoStop}%
\bibitem [{\citenamefont {Freitas}, \citenamefont {Proesmans},\ and\
  \citenamefont {Esposito}(2022)}]{Freitas2022reliability}%
  \BibitemOpen
  \bibfield  {author} {\bibinfo {author} {\bibfnamefont {N.}~\bibnamefont
  {Freitas}}, \bibinfo {author} {\bibfnamefont {K.}~\bibnamefont {Proesmans}},
  \ and\ \bibinfo {author} {\bibfnamefont {M.}~\bibnamefont {Esposito}},\
  }\href@noop {} {\bibfield  {journal} {\bibinfo  {journal} {Physical Review
  E}\ }\textbf {\bibinfo {volume} {105}},\ \bibinfo {pages} {034107} (\bibinfo
  {year} {2022})}\BibitemShut {NoStop}%
\bibitem [{\citenamefont {Van~Brandt}\ \emph {et~al.}(2023)\citenamefont
  {Van~Brandt}, \citenamefont {Silveira}, \citenamefont {Delvenne},\ and\
  \citenamefont {Flandre}}]{EUROSOI2023}%
  \BibitemOpen
  \bibfield  {author} {\bibinfo {author} {\bibfnamefont {L.}~\bibnamefont
  {Van~Brandt}}, \bibinfo {author} {\bibfnamefont {F.}~\bibnamefont
  {Silveira}}, \bibinfo {author} {\bibfnamefont {J.-C.}\ \bibnamefont
  {Delvenne}}, \ and\ \bibinfo {author} {\bibfnamefont {D.}~\bibnamefont
  {Flandre}},\ }in\ \href@noop {} {\emph {\bibinfo {booktitle} {9th Joint
  International EUROSOI Workshop and International Conference on Ultimate
  Integration on Silicon (EUROSOI-ULIS), Tarragona, Spain}}}\ (\bibinfo {year}
  {2023})\BibitemShut {NoStop}%
\bibitem [{\citenamefont {Landauer}(1961)}]{Landauer1961}%
  \BibitemOpen
  \bibfield  {author} {\bibinfo {author} {\bibfnamefont {R.}~\bibnamefont
  {Landauer}},\ }\href@noop {} {\bibfield  {journal} {\bibinfo  {journal} {IBM
  journal of research and development}\ }\textbf {\bibinfo {volume} {5}},\
  \bibinfo {pages} {183} (\bibinfo {year} {1961})}\BibitemShut {NoStop}%
\bibitem [{\citenamefont {Peliti}\ and\ \citenamefont
  {Pigolotti}(2021)}]{Peliti2021}%
  \BibitemOpen
  \bibfield  {author} {\bibinfo {author} {\bibfnamefont {L.}~\bibnamefont
  {Peliti}}\ and\ \bibinfo {author} {\bibfnamefont {S.}~\bibnamefont
  {Pigolotti}},\ }\href@noop {} {\emph {\bibinfo {title} {{Stochastic
  Thermodynamics: An Introduction}}}}\ (\bibinfo  {publisher} {Princeton
  University Press},\ \bibinfo {year} {2021})\BibitemShut {NoStop}%
\bibitem [{\citenamefont {Horowitz}\ and\ \citenamefont
  {Gingrich}(2020)}]{Horowitz2020}%
  \BibitemOpen
  \bibfield  {author} {\bibinfo {author} {\bibfnamefont {J.~M.}\ \bibnamefont
  {Horowitz}}\ and\ \bibinfo {author} {\bibfnamefont {T.~R.}\ \bibnamefont
  {Gingrich}},\ }\href@noop {} {\bibfield  {journal} {\bibinfo  {journal}
  {Nature Physics}\ }\textbf {\bibinfo {volume} {16}},\ \bibinfo {pages} {15}
  (\bibinfo {year} {2020})}\BibitemShut {NoStop}%
\bibitem [{\citenamefont {Falasco}, \citenamefont {Esposito},\ and\
  \citenamefont {Delvenne}(2020)}]{Falasco2020}%
  \BibitemOpen
  \bibfield  {author} {\bibinfo {author} {\bibfnamefont {G.}~\bibnamefont
  {Falasco}}, \bibinfo {author} {\bibfnamefont {M.}~\bibnamefont {Esposito}}, \
  and\ \bibinfo {author} {\bibfnamefont {J.-C.}\ \bibnamefont {Delvenne}},\
  }\href@noop {} {\bibfield  {journal} {\bibinfo  {journal} {New Journal of
  Physics}\ }\textbf {\bibinfo {volume} {22}},\ \bibinfo {pages} {053046}
  (\bibinfo {year} {2020})}\BibitemShut {NoStop}%
\bibitem [{\citenamefont {Shiraishi}, \citenamefont {Funo},\ and\ \citenamefont
  {Saito}(2018)}]{Shiraishi2018}%
  \BibitemOpen
  \bibfield  {author} {\bibinfo {author} {\bibfnamefont {N.}~\bibnamefont
  {Shiraishi}}, \bibinfo {author} {\bibfnamefont {K.}~\bibnamefont {Funo}}, \
  and\ \bibinfo {author} {\bibfnamefont {K.}~\bibnamefont {Saito}},\
  }\href@noop {} {\bibfield  {journal} {\bibinfo  {journal} {Physical review
  letters}\ }\textbf {\bibinfo {volume} {121}},\ \bibinfo {pages} {070601}
  (\bibinfo {year} {2018})}\BibitemShut {NoStop}%
\bibitem [{\citenamefont {Van~Brandt}\ and\ \citenamefont
  {Delvenne}(2023)}]{IEEENANO2023}%
  \BibitemOpen
  \bibfield  {author} {\bibinfo {author} {\bibfnamefont {L.}~\bibnamefont
  {Van~Brandt}}\ and\ \bibinfo {author} {\bibfnamefont {J.-C.}\ \bibnamefont
  {Delvenne}},\ }in\ \href@noop {} {\emph {\bibinfo {booktitle} {23rd IEEE
  International Conference on Nanotechnology (IEEE-NANO 2023)}}}\ (\bibinfo
  {year} {2023})\BibitemShut {NoStop}%
\bibitem [{\citenamefont {Kubo}(1966)}]{Kubo1966}%
  \BibitemOpen
  \bibfield  {author} {\bibinfo {author} {\bibfnamefont {R.}~\bibnamefont
  {Kubo}},\ }\href@noop {} {\bibfield  {journal} {\bibinfo  {journal} {Reports
  on progress in physics}\ }\textbf {\bibinfo {volume} {29}},\ \bibinfo {pages}
  {255} (\bibinfo {year} {1966})}\BibitemShut {NoStop}%
\bibitem [{\citenamefont {Johnson}(1928)}]{Johnson1928}%
  \BibitemOpen
  \bibfield  {author} {\bibinfo {author} {\bibfnamefont {J.~B.}\ \bibnamefont
  {Johnson}},\ }\href@noop {} {\bibfield  {journal} {\bibinfo  {journal}
  {Physical review}\ }\textbf {\bibinfo {volume} {32}},\ \bibinfo {pages} {97}
  (\bibinfo {year} {1928})}\BibitemShut {NoStop}%
\bibitem [{\citenamefont {Nyquist}(1928)}]{Nyquist1928}%
  \BibitemOpen
  \bibfield  {author} {\bibinfo {author} {\bibfnamefont {H.}~\bibnamefont
  {Nyquist}},\ }\href@noop {} {\bibfield  {journal} {\bibinfo  {journal}
  {Physical review}\ }\textbf {\bibinfo {volume} {32}},\ \bibinfo {pages} {110}
  (\bibinfo {year} {1928})}\BibitemShut {NoStop}%
\bibitem [{\citenamefont {Wyatt}\ and\ \citenamefont
  {Coram}(1999)}]{Wyatt1999}%
  \BibitemOpen
  \bibfield  {author} {\bibinfo {author} {\bibfnamefont {J.~L.}\ \bibnamefont
  {Wyatt}}\ and\ \bibinfo {author} {\bibfnamefont {G.~J.}\ \bibnamefont
  {Coram}},\ }\href@noop {} {\bibfield  {journal} {\bibinfo  {journal} {IEEE
  Transactions on Electron Devices}\ }\textbf {\bibinfo {volume} {46}},\
  \bibinfo {pages} {184} (\bibinfo {year} {1999})}\BibitemShut {NoStop}%
\bibitem [{\citenamefont {Jarzynski}(1997)}]{Jarzynski1997}%
  \BibitemOpen
  \bibfield  {author} {\bibinfo {author} {\bibfnamefont {C.}~\bibnamefont
  {Jarzynski}},\ }\href@noop {} {\bibfield  {journal} {\bibinfo  {journal}
  {Physical Review Letters}\ }\textbf {\bibinfo {volume} {78}},\ \bibinfo
  {pages} {2690} (\bibinfo {year} {1997})}\BibitemShut {NoStop}%
\bibitem [{\citenamefont {Delvenne}\ and\ \citenamefont
  {Van~Brandt}(2023)}]{IFAC2023}%
  \BibitemOpen
  \bibfield  {author} {\bibinfo {author} {\bibfnamefont {J.-C.}\ \bibnamefont
  {Delvenne}}\ and\ \bibinfo {author} {\bibfnamefont {L.}~\bibnamefont
  {Van~Brandt}},\ }in\ \href@noop {} {\emph {\bibinfo {booktitle} {22nd IFAC
  World Congress 2023, Yokohama, Japan, arXiv preprint arXiv:2208.14506}}}\
  (\bibinfo {year} {2023})\BibitemShut {NoStop}%
\bibitem [{\citenamefont {Falasco}, \citenamefont {Esposito},\ and\
  \citenamefont {Delvenne}(2022)}]{Falasco2022}%
  \BibitemOpen
  \bibfield  {author} {\bibinfo {author} {\bibfnamefont {G.}~\bibnamefont
  {Falasco}}, \bibinfo {author} {\bibfnamefont {M.}~\bibnamefont {Esposito}}, \
  and\ \bibinfo {author} {\bibfnamefont {J.-C.}\ \bibnamefont {Delvenne}},\
  }\href@noop {} {\bibfield  {journal} {\bibinfo  {journal} {Journal of Physics
  A: Mathematical and Theoretical}\ }\textbf {\bibinfo {volume} {55}},\
  \bibinfo {pages} {124002} (\bibinfo {year} {2022})}\BibitemShut {NoStop}%
\bibitem [{\citenamefont {Einstein}(1905)}]{Einstein1905}%
  \BibitemOpen
  \bibfield  {author} {\bibinfo {author} {\bibfnamefont {A.}~\bibnamefont
  {Einstein}},\ }\href@noop {} {\bibfield  {journal} {\bibinfo  {journal}
  {Investigations on the theory of the Brownian movement}\ } (\bibinfo {year}
  {1905})}\BibitemShut {NoStop}%
\bibitem [{\citenamefont {Baker}(2010)}]{Baker}%
  \BibitemOpen
  \bibfield  {author} {\bibinfo {author} {\bibfnamefont {R.~J.}\ \bibnamefont
  {Baker}},\ }\href@noop {} {\emph {\bibinfo {title} {{CMOS Circuit Design,
  Layout, and Simulation}}}},\ \bibinfo {edition} {3rd}\ ed.\ (\bibinfo
  {publisher} {John Wiley \& Sons, IEEE Press},\ \bibinfo {year}
  {2010})\BibitemShut {NoStop}%
\bibitem [{\citenamefont {Tsividis}\ and\ \citenamefont
  {McAndrew}(2011)}]{Tsividis2011}%
  \BibitemOpen
  \bibfield  {author} {\bibinfo {author} {\bibfnamefont {Y.}~\bibnamefont
  {Tsividis}}\ and\ \bibinfo {author} {\bibfnamefont {C.}~\bibnamefont
  {McAndrew}},\ }\href@noop {} {\emph {\bibinfo {title} {{Operation and
  Modeling of the MOS Transistor}}}},\ The Oxford Series in Electrical and
  Computer Engineering Series\ (\bibinfo  {publisher} {Oxford University
  Press},\ \bibinfo {year} {2011})\BibitemShut {NoStop}%
\bibitem [{\citenamefont {Poiroux}\ \emph {et~al.}(2013)\citenamefont
  {Poiroux}, \citenamefont {Rozeau}, \citenamefont {Martinie}, \citenamefont
  {Scheer}, \citenamefont {Puget}, \citenamefont {Jaud}, \citenamefont
  {El~Ghouli}, \citenamefont {Barb{\'e}}, \citenamefont {Juge},\ and\
  \citenamefont {Faynot}}]{Poiroux2013}%
  \BibitemOpen
  \bibfield  {author} {\bibinfo {author} {\bibfnamefont {T.}~\bibnamefont
  {Poiroux}}, \bibinfo {author} {\bibfnamefont {O.}~\bibnamefont {Rozeau}},
  \bibinfo {author} {\bibfnamefont {S.}~\bibnamefont {Martinie}}, \bibinfo
  {author} {\bibfnamefont {P.}~\bibnamefont {Scheer}}, \bibinfo {author}
  {\bibfnamefont {S.}~\bibnamefont {Puget}}, \bibinfo {author} {\bibfnamefont
  {M.}~\bibnamefont {Jaud}}, \bibinfo {author} {\bibfnamefont {S.}~\bibnamefont
  {El~Ghouli}}, \bibinfo {author} {\bibfnamefont {J.}~\bibnamefont
  {Barb{\'e}}}, \bibinfo {author} {\bibfnamefont {A.}~\bibnamefont {Juge}}, \
  and\ \bibinfo {author} {\bibfnamefont {O.}~\bibnamefont {Faynot}},\ }in\
  \href@noop {} {\emph {\bibinfo {booktitle} {2013 IEEE International Electron
  Devices Meeting}}}\ (\bibinfo {organization} {IEEE},\ \bibinfo {year}
  {2013})\ pp.\ \bibinfo {pages} {p.12.4.1--12.4.4}\BibitemShut {NoStop}%
\bibitem [{\citenamefont {Papoulis}(1991)}]{Papoulis1991}%
  \BibitemOpen
  \bibfield  {author} {\bibinfo {author} {\bibfnamefont {A.}~\bibnamefont
  {Papoulis}},\ }\href@noop {} {\emph {\bibinfo {title} {Probability, Random
  Variables, and Stochastic Processes}}},\ McGraw-Hill Series in Electrical
  Engineering\ (\bibinfo  {publisher} {McGraw-Hill},\ \bibinfo {year}
  {1991})\BibitemShut {NoStop}%
\bibitem [{\citenamefont {Snider}\ \emph {et~al.}(2012)\citenamefont {Snider},
  \citenamefont {Blair}, \citenamefont {Thorpe}, \citenamefont {Appleton},
  \citenamefont {Boechler}, \citenamefont {Orlov},\ and\ \citenamefont
  {Lent}}]{Snider2012}%
  \BibitemOpen
  \bibfield  {author} {\bibinfo {author} {\bibfnamefont {G.~L.}\ \bibnamefont
  {Snider}}, \bibinfo {author} {\bibfnamefont {E.~P.}\ \bibnamefont {Blair}},
  \bibinfo {author} {\bibfnamefont {C.~C.}\ \bibnamefont {Thorpe}}, \bibinfo
  {author} {\bibfnamefont {B.~T.}\ \bibnamefont {Appleton}}, \bibinfo {author}
  {\bibfnamefont {G.~P.}\ \bibnamefont {Boechler}}, \bibinfo {author}
  {\bibfnamefont {A.~O.}\ \bibnamefont {Orlov}}, \ and\ \bibinfo {author}
  {\bibfnamefont {C.~S.}\ \bibnamefont {Lent}},\ }in\ \href@noop {} {\emph
  {\bibinfo {booktitle} {2012 12th IEEE International Conference on
  Nanotechnology (IEEE-NANO)}}}\ (\bibinfo {organization} {IEEE},\ \bibinfo
  {year} {2012})\ pp.\ \bibinfo {pages} {1--6}\BibitemShut {NoStop}%
\bibitem [{\citenamefont {Orlov}\ \emph {et~al.}(2012)\citenamefont {Orlov},
  \citenamefont {Lent}, \citenamefont {Thorpe}, \citenamefont {Boechler},\ and\
  \citenamefont {Snider}}]{Orlov2012}%
  \BibitemOpen
  \bibfield  {author} {\bibinfo {author} {\bibfnamefont {A.~O.}\ \bibnamefont
  {Orlov}}, \bibinfo {author} {\bibfnamefont {C.~S.}\ \bibnamefont {Lent}},
  \bibinfo {author} {\bibfnamefont {C.~C.}\ \bibnamefont {Thorpe}}, \bibinfo
  {author} {\bibfnamefont {G.~P.}\ \bibnamefont {Boechler}}, \ and\ \bibinfo
  {author} {\bibfnamefont {G.~L.}\ \bibnamefont {Snider}},\ }\href@noop {}
  {\bibfield  {journal} {\bibinfo  {journal} {Japanese Journal of Applied
  Physics}\ }\textbf {\bibinfo {volume} {51}},\ \bibinfo {pages} {06FE10}
  (\bibinfo {year} {2012})}\BibitemShut {NoStop}%
\end{thebibliography}%

\end{document}